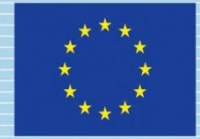
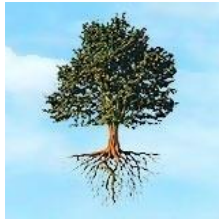
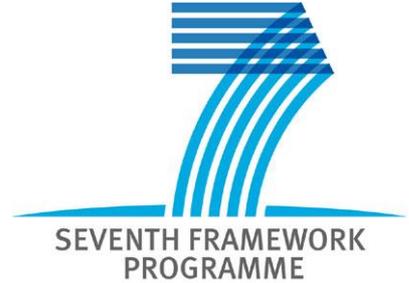

**SEVENTH FRAMEWORK PROGRAMME**
Call FP7-ICT-2009-4    Objective ICT-2009.8.1
**[FET Proactive 1: Concurrent Tera-device Computing]**

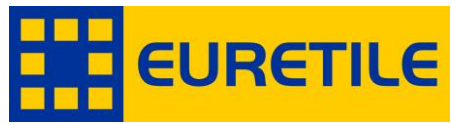

Project acronym: **EURETILE**
Project full title: **European Reference Tiled Architecture Experiment**
Grant agreement no.: **247846**

# EURETILE D7.3 – Dynamic DAL benchmark coding, measurements on MPI version of DPSNN-STDP (distributed plastic spiking neural net) and improvements to other DAL codes

**Lead contractor for deliverable: INFN**
Due date of submission: 2014-02-28

**Revision:** See document footer.

| Project co-funded by the European Commission within the Seventh Framework Programme ||
|---|---|
| **Dissemination Level: PU** ||
| **PU** | Public |
| **PP** | Restricted to other programme participants (including the Commission Services) |
| **RE** | Restricted to a group specified by the consortium (including the Commission Services) |
| **CO** | Confidential, only for members of the consortium (including the Commission Services) |

Project: **EURETILE** – European Reference Tiled Architecture Experiment
Grant Agreement no.: **247846**
Call: FP7-ICT-2009-4 Objective: FET - ICT-2009.8.1 Concurrent Tera-device Computing

# Dynamic DAL benchmark coding, measurements on MPI version of DPSNN-STDP (distributed plastic spiking neural net) and improvements to other DAL codes

*Pier Stanislao Paolucci[1], Iuliana Bacivarov[2], Devendra Rai[2], Lars Schor[2], Lothar Thiele[2], Hoeseok Yang[2], Elena Pastorelli[1], Roberto Ammendola[3], Andrea Biagioni[1], Ottorino Frezza[1], Francesca Lo Cicero[1], Alessandro Lonardo[1], Francesco Simula[1], Laura Tosoratto[1], Piero Vicini[1]*

**[1]INFN Roma "Sapienza"**
**[2]Computer Engineering and Networks Laboratory, ETH Zurich**
**[3]INFN Tor Vergata**

Authorship notes:

- Corresponding author: Pier Stanislao Paolucci: e-mail: pier.paolucci@roma1.infn.it
- The team of ETH Zurich is author of the "Advanced Benchmarks for Embedded Systems" Chapter.
- The team of INFN is author of the "Brain Simulation Benchmark" Chapter;

Project: **EURETILE** – European Reference Tiled Architecture Experiment
Grant Agreement no.: **247846**
Call: FP7-ICT-2009-4 Objective: FET - ICT-2009.8.1 Concurrent Tera-device Computing

# Table of Contents

# 1. Introduction

The EURETILE project is about the software and hardware architecture of future many-tile parallel/distributed fault-tolerant systems. We focus on dynamic workloads characterized by heavy numerical processing requirements. The ambition is to identify common techniques that could be applied to both the Embedded Systems and HPC domains. EURETILE required the selection and coding of a set of dedicated benchmarks.

A first set of benchmarks includes ray-tracing, the H.264 codec, a recursive array-sorting, a picture-in-picture (PiP) video decoder, plus a few classical DSP/Data-flow kernels, like fast discrete Fourier transform (FFT), N-order infinite impulse response filter (IIR) and matrix multiplication. Several of these applications originated in the HPC domain, but will migrate on next generation mobile computing systems, equipped with powerful many-core many-tile processors.

A second domain selected by EURETILE is the simulation of the activity and plasticity of large networks of neurons and synapses. This kind of simulations constitutes an extreme challenge for High Performance Computing systems, but at smaller scales could be applied to embedded systems, if included in robots and medical systems. Indeed, the human brain learns through the dynamic of more than $10^{15}$ synapses and involves in its computations approximately $10^{11}$ neurons. The brain consumes a few tens of Watts; the problem is, even at the highest levels of abstraction, a whole brain simulation would require tens of Giga Watts on computing systems designed using current technologies. Fault-tolerance and scaling to the required computational power are also overwhelming on conventional supercomputing architectures.

This document (D7.3) is the first public deliverable of EURETILE Work Package 7 "Challenging Tiled Applications". The goal of WP7 is to define and code a set of benchmarks for the EURETILE programming environment and for the EURETILE execution platforms. It is interesting to compare the features (and bottlenecks) of consolidated environments versus those conceived by EURETILE. Therefore, we decided that a subset of the benchmarks had to be coded also for easy porting and compilation on both the EURETILE environment as well as on standard distributed/parallel programming environments, like MPI and OpenMP.

During 2013, WP7 had two main goals: 1- the coding of a set of benchmarks that show a dynamic behavior, coded using the DAL programming environment; 2- the finalization of an efficient and scalable version of the cortical simulation benchmark (DPSNN-STDP) in MPI, and the measurement of its performance on a QUonG platform, posing the ground for the final porting and measurements on the EURETILE environment, to be performed during 2014.

We present results produced during 2013. However, as WP7 produced two non-public deliverables (D7.1. and D7.2) describing 2011 and 2012 activities, short summaries of results produced during earlier stages of the project have been inserted to create a self-consistent document.

The integration of the EURETILE software environment is the goal of the Work Package 8 "SW Tool-chain integration". A basic version of the DPSNN-STDP code has been ported during 2013 to the DAL programming environment for execution on the EURETILE hardware environment, supported by the DNA-OS operating system. The description of the specificities of the DPSNN-STDP DAL version used for testing the 2013 release of the EURETILE environment are described in the document D8.2 "Inter-tile and multi-core tool chains for complex applications". The D8.2 document includes also a summary description of other DAL codes used to test the 2013 release of the full environment.

Work Package 6 "Innovations on HW Intellectual Properties" addresses the experimental EURETILE hardware execution platform; WP5 "Many-tile Simulation/Profiling" includes the creation of the simulated execution platform. Therefore, a description of the aspects concerning the execution and profiling of the benchmarks coded by WP7 on the EURETILE platforms can be found in the set of deliverable documents D6.x and D5.x.

The development of the DAL programming environment is the main goal of WP3 "Foundational Many-Process Programming Environment", while the generation of the DNA OS services for the heterogeneous multicore system is the main objective of WP4 "Distributed Hardware Dependent Software Generation". Therefore, the reader is addressed to the set of deliverables D3.x and D4.x for a more complete description of the DAL programming environment and of the DNA-OS operating system.

## 2. Advanced Benchmarks for Embedded Systems

This chapter summarizes the progress of programming benchmarks for embedded systems in the context of EURETILE WP7 "Challenging Tiled Application" with a strong focus on the progress achieved during the year 2013. The distributed application layer (DAL) formalism will be used to describe the benchmarks. DAL is a scenario-based design flow that supports the design, optimization, and simultaneous execution of multiple applications targeting heterogeneous many-core systems. We refer the reader to [1, 2, 3, 4, 5, 6] in Section 2.5 "References" for more details. In particular, an overview of the design flow is given in [1]. The mapping optimization framework is summarized in [2] and thermal-aware constraints are integrated into the optimization framework in [3, 4]. The efficient execution of applications specified using the DAL formalism is described in [5] targeting distributed memory systems and in [6] targeting heterogeneous multi-core systems using OpenCL.

This document (D7.3) is the first public deliverable of WP7. WP7 produced two non-public deliverables (D7.1. and D7.2) describing 2011 and 2012 activities. Therefore, the chapter starts with Section 2.1 that summarizes various DSP/data-flow kernels developed earlier in the project. Afterwards, Sections 2.2, 2.3, and 2.4 describe benchmarks that have been developed in 2013, in conjunction with the semantics of expandable process networks (EPNs) that are proposed in [7] and developed in the framework of work-package 3. EPNs extend conventional streaming programming models by abstracting several possible granularities in a single specification. This enables the application to vary its degree of parallelism if the computational demand of an application changes or if the number of available PEs varies due to applications that are started or stopped.

In particular, the first described benchmark is a ray-tracing algorithm. Providing a high degree of realism, ray tracing is expected to be implemented as a real-time rendering algorithm in the next generation of embedded many-core systems. Afterwards, we describe a distributed sorting algorithm. The application uses the quicksort algorithm whereby the sorting kernel can be replaced with a structural description increasing the available degree of parallelism. Finally, we summarize our work to explore intra-application dynamism using an H.264 standard complying codec pair.

## 2.1 Summary of DSP/Data-flow Kernels

We start with an overview of the work performed in 2011 and 2012. Sections 2.2, 2.3, and 2.4 describe new benchmarks developed during 2013. To this end, we summarize three applications that are typically used in digital signal processing (DSP), namely a distributed implementation of an *N*-point discrete Fourier transformation (DFT), an *N*-order IIR filter, and a distributed implementation of a matrix multiplication. Afterwards, to demonstrate the expected dynamism in future embedded systems and how this dynamism can be described by the DAL formalism, an advanced picture-in-picture (PiP) software for embedded video processing systems is described. The described application examples have been added as programming examples in the graphical software development environment DALipse that extends Eclipse with the ability to visually specify DAL applications and their interactions [8]. More details about the benchmarks are given in [9], which describes the technical implementation of the benchmarks.

### 2.1.1 Discrete Fourier Transformation (FFT)

The first benchmark application is an *N*-point discrete Fourier transform (DFT). The DFT is a block operation, which takes *N* complex-valued (time-domain) input coefficients and transforms them into *N* complex-valued (frequency-domain) output coefficients. An efficient implementation of the DFT is the decimation-in-time radix-2 FFT. In this implementation, the computation is split up into $N/2 \cdot \log_2(N)$ 2-point FFTs. The 2-point FFT is shown in Figure 2-1. Using the 2-point FFT, higher-order FFTs can be computed by arranging $N/2 \cdot \log_2(N)$ 2-point FFTs and connecting them using a butterfly network.

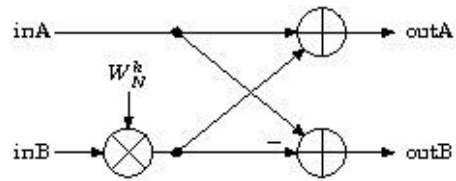

Figure 2-1. 2-point FFT.

### 2.1.2 N-Order Infinite Impulse Response Filter (IIR)

The *N*-order infinite impulse response filter (IIR) can be described by the following state equation:

$$\boldsymbol{y}[n] = \frac{1}{a_0}(\boldsymbol{b_0} \cdot \boldsymbol{x}[n] + \boldsymbol{b_1} \cdot \boldsymbol{x}[n-1] + \cdots + \boldsymbol{b_P} \cdot \boldsymbol{x}[n-P]$$
$$- a_1 \cdot \boldsymbol{y}[n-1] - a_2 \cdot \boldsymbol{y}[n-2] - \cdots - a_Q \cdot \boldsymbol{y}[n-Q])$$

with *x* the sequence of input samples and *y* the sequence of output samples. A possible implementation of an *N*-order IIR filter is a decomposition of first order IIR filters as shown in Figure 2-2.

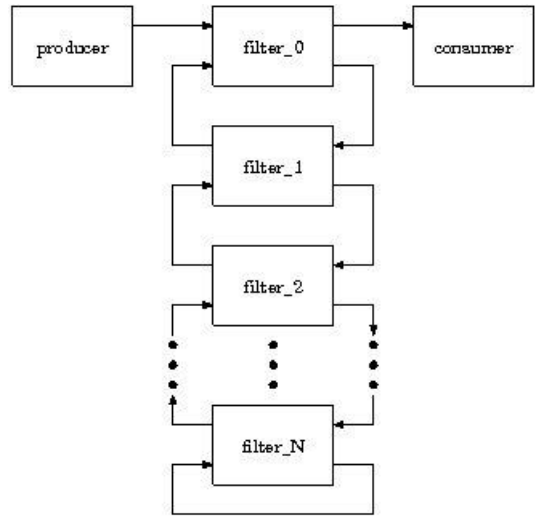

Figure 2-2. Process network of *N*-th order IIR filter.

### 2.1.3 Matrix Multiplication

Next, we consider the *NxN* matrix multiplication. It splits the computation of the matrix product up into single multiplications and additions. This way, the application "matrix multiplication" is split up into single processes, each of which only performs a multiplication followed by an addition. Single coefficients of the resulting matrix are computed by appropriately connecting the processes.

### 2.1.4 Picture-In-Picture (PiP) Video Decoder

Last, we summarize a picture-in-picture (PiP) software for embedded video processing systems that demonstrates the capabilities of the reliable formalism proposed in [1] to describe and manage dynamic scenarios.

Figure 2-3 shows a screenshot of DALipse with the finite state machine (FSM) of the considered PiP software. The software is composed of eight scenarios and three different video decoder applications. The *HD* application processes high-definition, the *SD* application standard-definition, and the *VCD* application low-resolution video data. The software has two major execution modes, namely watching high-definition (scenario *HD*) or standard-definition videos (scenario *SD*). In addition, the user might want to pause the video or watch a preview of another video by activating the PiP mode (i.e., starting the *VCD* application). Due to resource restrictions, the user is only able to activate the PiP mode when the *SD* application is running or paused, or the *HD* application is paused.

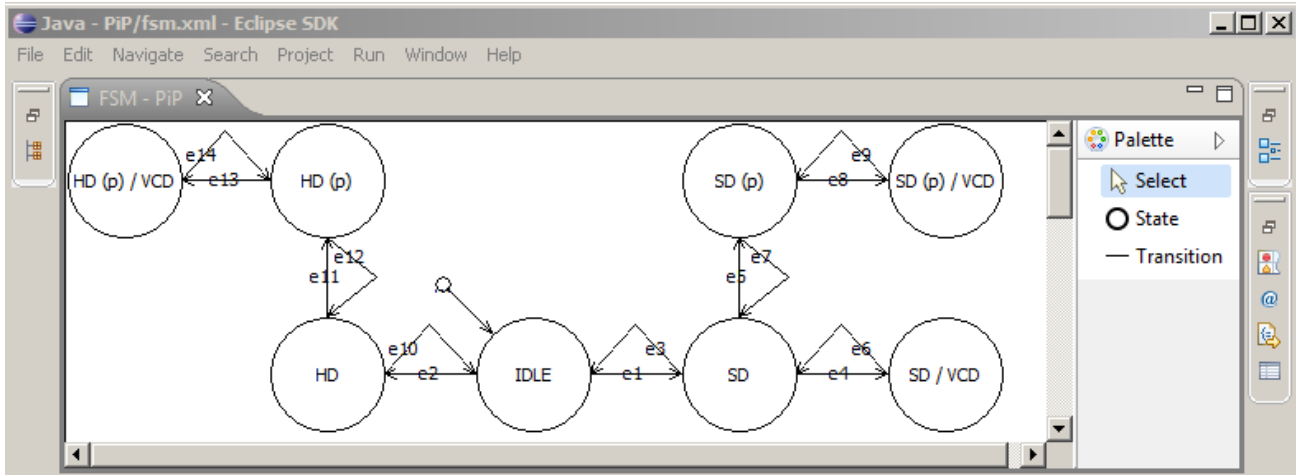

Figure 2-3. Finite state machine of the PiP software. Paused applications are indicated by a (p)

## 2.2 Ray Tracing

Ray tracing is often praised for its very high degree of realism, made possible by the algorithm's close similarity to how vision and light works in nature. As it has high computational cost, ray tracing is only rarely used on the current embedded multi-core platforms. However, the next generation of embedded many-core systems is supposed to provide a tremendously increased computational performance enabling the use of real-time rendering algorithms like ray tracing.

### 2.2.1 Benchmark Description

Ray tracing is a technique used in computer graphics for generating (also called rendering) an image from a scene description [10]. The scene description contains information on all objects present in the scene and is usually implemented as a list consisting of geometric objects. A scene is generally an approximation of the reality and can be described using various different geometric primitives. In addition to all objects present in the scene, the scene description also contains information on the camera position, direction, and viewing angle.

With the information contained in the scene description, an application is now able to render an image of the scene. This can be thought of as corresponding to the act of taking an actual photograph in real life. The algorithm we are using to render the image is called path tracing. It is very similar to ray tracing, with a slight adjustment, which has the advantage of being able to generate more advanced optical effects, such as caustics and soft shadows.

Ray tracing applications naturally consist of three logic parts. The first part of the application comprises the generation of the rays. The next part is the second and most computational intensive part of the application, the intersection of the rays with the scene, which is to be rendered. Lastly, the calculated values are aggregated and stored to an image file in the third part. We decided to use this partition for the process network specification, as it allows us to neatly separate the intersection part, which we are interested in parallelizing as it has the biggest contribution to the application's total execution time.

The resulting layout of the process network is illustrated in Figure 2-4. In the resulting implementation, the Generator process calculates all rays and sends them to the Intersect processes, which is where the actual path tracing takes place by calling the process's radiance method, which recursively calls itself until either no object was intersected or the maximum depth has been reached.

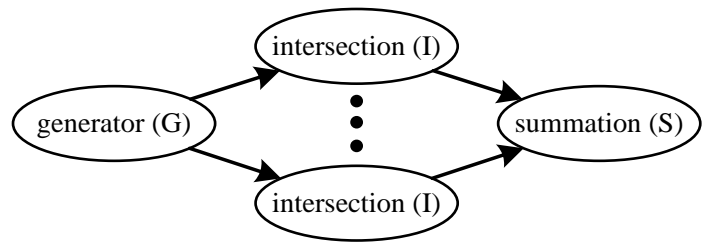

Figure 2-4. EPN specification of the ray-tracing application that support replication.

Various extensions of the basic process network have been considered to improve the performance of the ray tracing application on systems with more than three cores. First, multiple instances of the Intersect process might calculate different paths in parallel. Thus, the Generator process alternates the output port between rays, in order for all replicated Intersect processes to receive the rays. This is also shown in the EPN specification illustrated in Figure 2-4.

In order to obtain additional parallelism, we use another property of the ray-tracing algorithm. When a ray hits a surface, which is both refractive and reflective, two new rays are generated. Instead of recursively calling the function radiance, two completely new Intersect processes, which will continue the computation from where the first Intersect process has stopped its computations. In order for the new processes to do this, the computed rays are sent to them over the communication channel together with the values of all the local variables, which are needed to complete the remainder of the computation. These variables are the recursion depth, the intersection's reflectivity / transmission coefficient and the intersected object's color.

### 2.2.2 Comparison with OpenMP

After implementing all these extensions, we compared an OpenMP implementation of the ray tracing application to the DAL implementation. As a test machine, we used an ordinary Intel quad-core processor. When creating a 100×100 pixel output image of the Cornell Box scene, a typical benchmark scene, at 100 samples per pixel, the original implementation using OpenMP completed in 25 seconds. In comparison to this, the DAL implementation running on the same machine finished the computation of the same scene after 27 seconds, which is only 8% slower. Note that the OpenMP implementation has been highly optimized, while the DAL framework was never optimized for maximum performance with this kind of application.

### 2.2.3 Speed-Up due to Parallelism

Finally, we evaluated the speed-up due to the available number of cores for the ray-tracing application. The ray-tracing algorithm generates again an image of 100 ×100 pixels and can concurrently analyze multiple rays. We mapped either one or two such processes onto one core. In Figure 2-5, the speed-up is compared for implementations running on a different number of cores of Intel's SCC processor [11]. The speed-up is calculated with respect to an implementation running on a single core. The SCC processor is a 48-core experimental processor from Intel. It consists of 24 tiles that are organized into a 4×6 grid and linked by a 2D mesh on-chip network. A tile contains a pair of P54C cores, a router, and a 16KB block of SRAM. Each core has an independent L1 and L2 cache. Overall, the ray-tracing algorithm achieves a speed-up of almost 20 on 24 cores. In particular, we can see that the ray-tracing algorithm is well suited for parallelization as each ray can individually be analyzed.

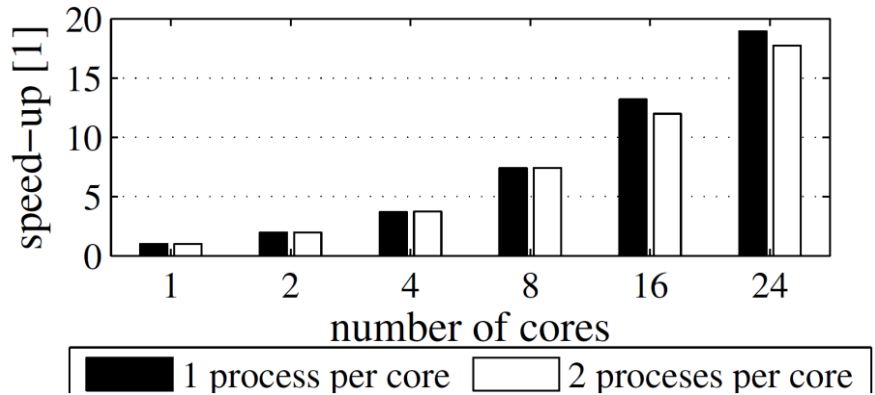

Figure 2-5. Speed-ups of the ray-tracing algorithm for a varying number of cores.

## 2.3 Recursive Array-Sorting

Next, we describe and evaluate the performance of a recursive array-sorting algorithm. Note that the conventional specification of process networks does not support such recursive applications; however, the proposed semantics of EPNs extend such conventional specifications so that recursive algorithms can be supported.

### 2.3.1 Benchmark Description

We particularly consider a quicksort algorithm. The EPN specification is illustrated in Figure 2-6 and Figure 2-7. Process "src" ("dest") generates (displays) the input (output) array, and process "sort" sorts the elements in ascending order. As the quicksort algorithm recursively sorts the array, process "sort" can be replaced by a structural description. "div" first partitions the array into two groups: the first group contains the elements that are smaller than the median value and the second group contains the remaining elements. The divided arrays are individually sent to a different instance of the "sort" process. Finally, the individually sorted sub-arrays are merged into a single array. Thus, by recursively unfolding the "sort" process, the original topology can be transformed to have more tasks. As the length of the array that each "sort" process has to sort is halved in each recursion step, the execution time is reduced with each recursion step, as well.

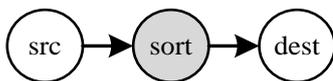

Figure 2-6. Top-level process network of the quicksort algorithm.

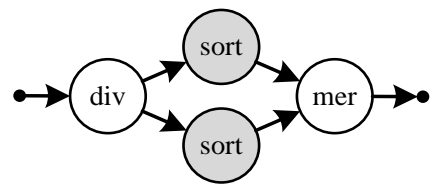

Figure 2-7. Structural description of the process "sort".

### 2.3.2 Speed-Up due to Parallelism

In Figure 2-8, the execution time to sort 5000 random arrays with each 5000 elements is compared for a varying number of available cores and different recursion depths. The reported numbers are obtained from running the final implementation on Intel's SCC processor that has been previously described. *No unfolding* refers to the basic quicksort algorithm illustrated in Figure 2-6. *X-times unfolded* refers to an implementation where the "sort" process is x times recursively unfolded. Finally, the EPN implementation leaves the task of unfolding to the optimization framework described in deliverable 3.3. The evaluation shows that the optimization framework selects a different unfolding degree depending on the available number of cores. On the one hand, if the number of cores is small, a low recursion depth achieves best performance as the switching and communication overhead is low. On the other hand, if the number of cores is large, a high recursion depth accomplishes a lower execution time as the array is sorted in parallel. Even though the EPN specification tries to optimize both the degree of parallelism and the mapping to minimize the execution time, the speed-up is much lower than the optimal speed-up.

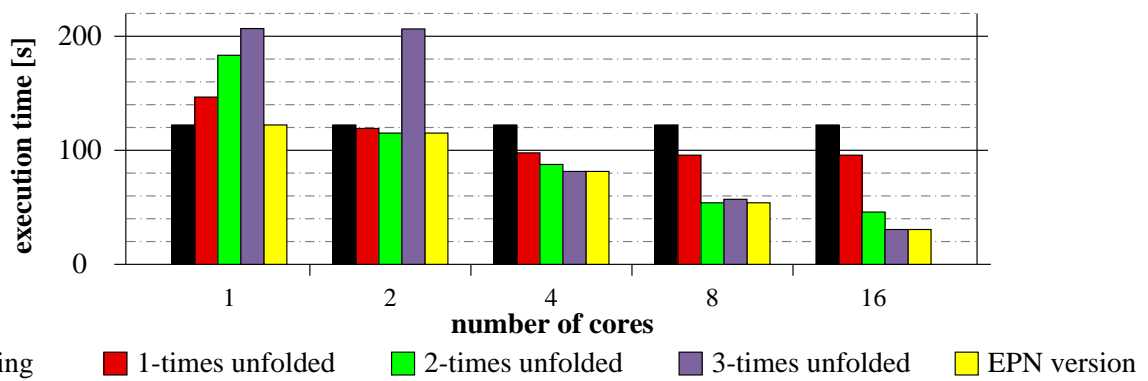

Figure 2-8. Execution time of quicksort for a varying number of available cores.

The speed-up is 2.3 when using 8 cores instead of 1 core and 4.1 when using sixteen cores instead of one core. This might be because additional time is spent to first partition the array into two groups and then to transmit the intermediate results between the different cores.

## 2.4 H.264 Codec

The H.264/MPEG-4 AVC (Advanced Video Codec) is one of the most widely used video coding standards in recent years, which is mainly attributed to its significantly increased achievable bitrate compression compared to its predecessors. However, the increased performance comes at the cost of increased complexity and a lengthy specification.

### 2.4.1 The H.264 Standard

The H.264 standard specification [12] was developed by the ISO/IEC Moving Picture Experts Group (MPEG) and the ITU-T Video Coding Experts Group (VCEG), a partnership known as Joint Video Team (JVT). The first version of the standard was completed in 2003 under the formal name ISO/IEC 1449-10 - MPEG 4 Part 10, Advanced Video Coding. It was motivated by emerging networking technologies such as xDSL and UMTS, and was designed to outperform previous video coding standards, such as the basic MPEG-4 (part 2 of the MPEG4 standards) and H263, providing better compression rates and fault-tolerance. The standard reportedly achieves up to 50% better compression in various bitrates and resolutions, although its decoder is twice as complex as the basic MPEG-4 and its encoder can be as much as 10 times more complex [13].

The standard provides flexibility and is applicable in a variety of cases, ranging from two-way video communication to high quality video streaming over packet networks. This flexibility is achieved by the different profiles that are described in the specification including:
- the Baseline profile;
- the Main profile; and
- the Extended profile.

Each profile supports a particular set of coding functions and targets different applications, but is not limited by them. In this work, we focus on the baseline profile, which has the lowest complexity and is more suited for embedded systems.

The standard is implemented via a pair of algorithms called codec (enCOder/DECoder). The encoder takes a video sequence as input and processes it to produce the compressed bit stream, which can be transmitted via some communication medium or stored. The decoder reverses the compression and reconstructs the video sequence for playback.

A video consists of a sequence of frames that, when alternated at the correct rate, produce the moving picture. The frame consists of spatio-temporal samples called picture elements or pixels. The H.264 standard supports rectangular frames of the YUV color space (a.k.a. YCrCb) sampled with the 4:2:0 format. As such, each picture is described by its luminance and Cr and Cb chroma coefficients (Y, U, V, respectively), with the last two being sampled at half the rate of the first, both in the horizontal and vertical directions.

A frame, in order to be encoded and to produce a coded picture, is divided into macroblocks, each containing 16x16 luma samples, 8x8 Cr chroma samples and 8x8 Cb chroma samples. A macroblock can possibly be further divided into sizes as small as 4x4 Y, 2x2 U and 2x2 V samples, if needed. A slice is a set of macroblocks in raster scan order that belong in the same picture and can be encoded independently to other slices of the same frame. For simplicity, we will only use one slice per frame. A slice can be characterized as an I- or P-slice in the baseline profile. An I-slice contains only I macroblocks, that is, macroblocks that are encoded (decoded) by utilizing data from previously encoded (decoded) macroblocks within the same slice and as such being independent from any other slice. A P-slice, on the other hand, contains both I and P macroblocks. P macroblocks are macroblocks that are encoded (decoded) by utilizing data from macroblocks that belong to a previously encoded (decoded) frame (called the reference frame). The process that enables the coding based on a previous frame is called Motion Estimation and Motion Compensation.

### 2.4.2 H.264 Structure

In common with earlier video coding standards, the H.264 specification does not directly specify a codec pair. On the contrary, it specifies in detail the syntax of a coded bitstream, the semantics of the syntax elements and their decoding process, by which the initial video sequence can be retrieved. In practice however, the functional representations of an H.264 encoder and decoder pair most likely resemble the ones shown in Figure 2-9 and Figure 2-10, respectively.

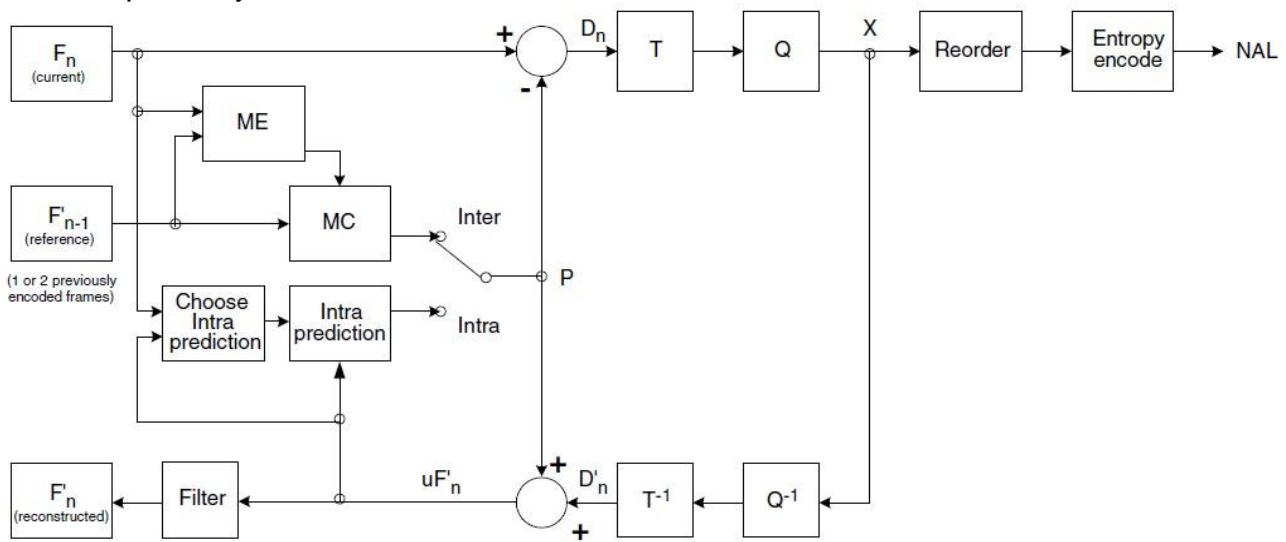

Figure 2-9. Functional specification of an H.264 encoder [14].

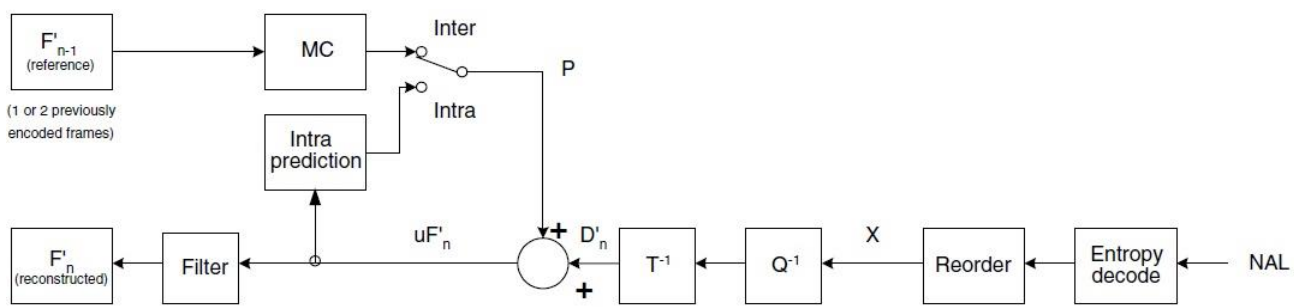

Figure 2-10. Functional specification of an H.264 decoder [14].

### 2.4.3 DAL Implementation of the Standard

Using the basic specification presented above, the base implementation for the encoder and decoder application on the DAL framework was created. The implementation is based on code for HOPES [16], which has been provided by the Seoul National University and supports the baseline profile of the coding standard. For brevity, we only describe the basic implementation of the encoder. For more information, we refer to [16]. The encoder implementation uses task division on the macroblock level. The set of functions is divided among five processes: *Init*, *ME*, *Encode*, *Deblock*, and *VLC*. The division of the application's function set is shown in Figure 2-11.

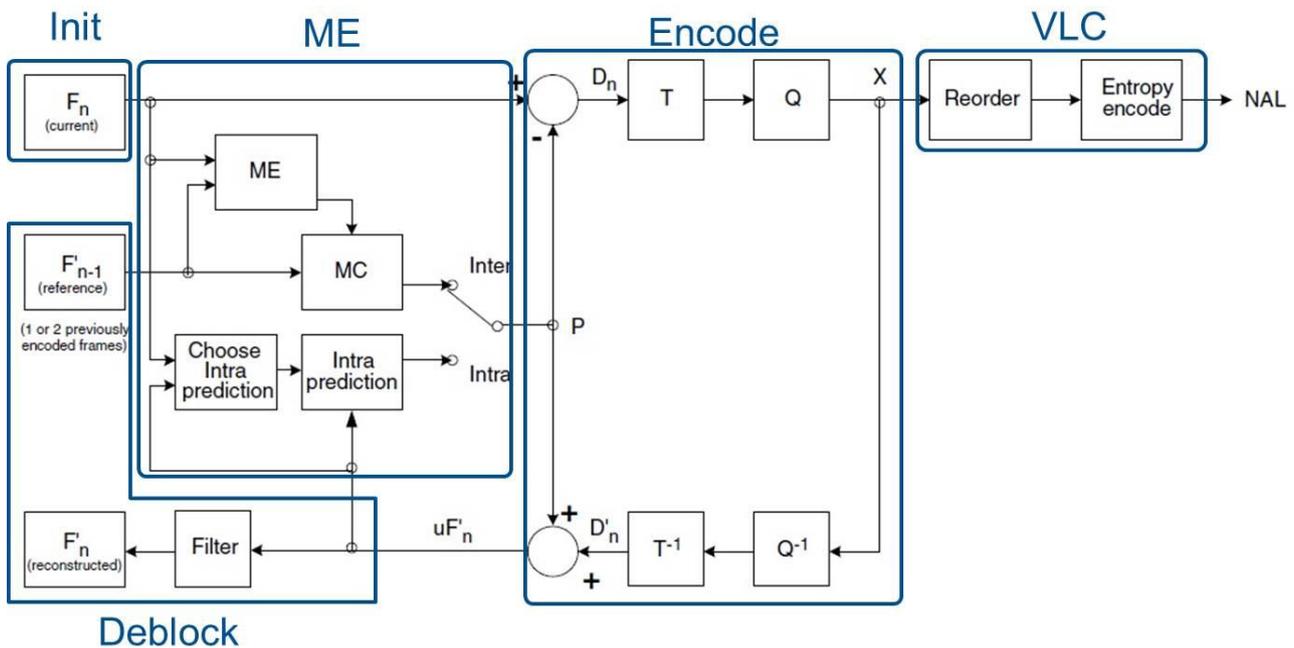

Figure 2-11. Division of the encoder functions in processes.

The ME process reads one macroblock in each firing sent from the Init process and performs inter- and intra-prediction, using the macroblock's reference and neighboring data that is communicated from the Deblock process. The results of the two predictions are compared and the best one is used as the macroblock's predictor, which is sent to the Encode process.

In turn, the Encode process uses the predictor and the raw macroblock to calculate the residual, which is subsequently transformed and quantized to form the coded macroblock. This result is sent to the VLC process and is reconstructed, with the reconstruction being passed onto the Deblock process.

The Deblock process stores the reconstructed macroblocks to form a database containing the reference frame's macroblocks used for inter-prediction, and the current frame's macroblocks used for intra-prediction. Thus, when it receives a reconstructed macroblock, it saves it in the database and loads macroblocks that are needed for inter- and intra-prediction of the next macroblock in line. The Deblock process also performs the task of deblocking filtering. Since intra-prediction requires unfiltered macroblocks, and inter-prediction requires the filtered ones, frames that are used as references are filtered when they are fully processed. When the final macroblock of such a frame is received, the process applies the deblocking filter to the whole frame and saves it as a reference.

Finally, the VLC process reorders the elements of the received coded macroblock and produces the final bitstream by applying variable length coding on these elements and every other necessary information needed for reconstructing the macroblock (e.g., the macroblock type, possibly the motion vector etc.). It also creates the headers of the NAL units and writes all this data on the output file.

### 2.4.4  Preliminary Evaluation Results

Finally, we describe some preliminary evaluation results when executing the H.264 codec on top of a dual 8-core Xeon E5-2690 CPU with 16 GB of shared memory. Two input video resolutions where used, QCIF (resolution: 176×144) and CIF (352×288).

**2.4.4.1 I to P Frames Ratio**

First, we study the effect of the ratio between I and P frames on the performance of the encoder. Since the first frame of each GOP is specified as an I frame and the rest as P frames, the ratio can be adjusted by increasing or decreasing the GOP size. Macroblocks belonging to I frames are only intra-predicted, while macroblocks from P frames are both inter- and intra-predicted. Thus, I frames are encoded faster than P frames, which means that higher fps can be achieved. However, the compression rate of P frames is higher, since both prediction techniques are used on their macroblocks. The more accurate one of the two predictions is selected, which in turn produces a residual with less energy, needing less bits to be encoded.

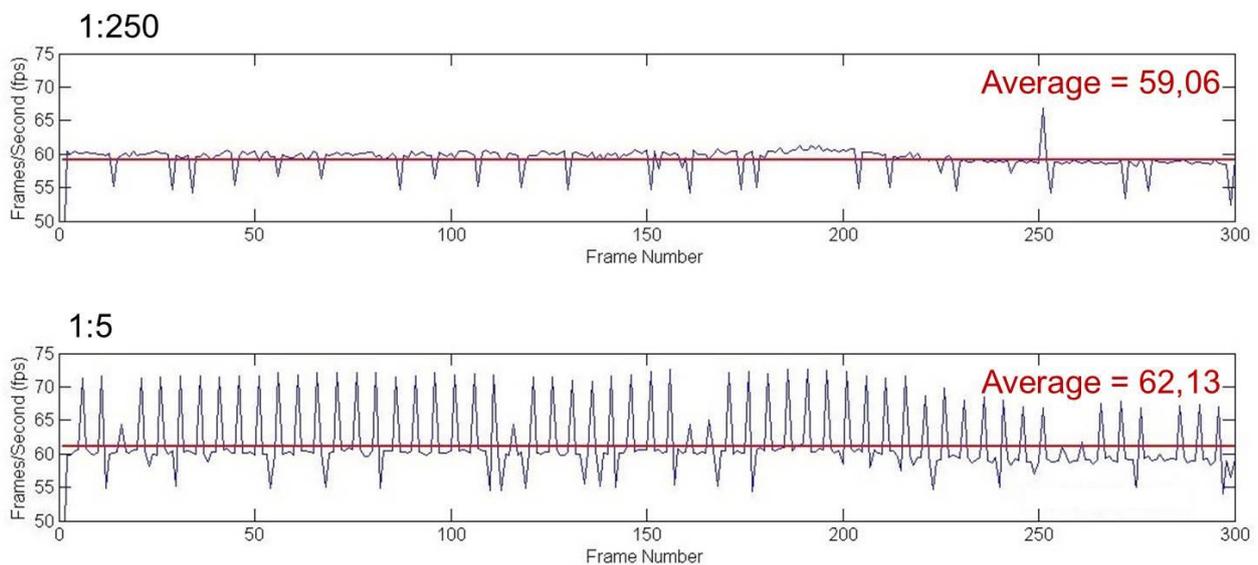

Figure 2-12. Fps per frame for 1:250 (top) and 1:5 (bottom) I to P frames encoding of a 300 frames, CIF resolution video.

Figure 2-12 shows the frames per second (fps) achieved for each frame for encoding 300 frames of a CIF resolution video using the implemented encoder and two different I to P frames ratios. For the 1:250 rate, it is evident that the fps rate for the P-frames is just below 60, but spikes on the sole I-frame (frame 251) to approximately 70 fps. By increasing the rate to 1:5, more spikes are introduced, which effectively increases the average fps of the encoding process by approximately 5.1%. The input 45.5 MB video is compressed to 3.82 MB for the lower ratio and to 4.04 MB for the higher ratio. This means that along with the speedup, a drop of approximately 5.4% is witnessed in the compression rate, as expected. These results show that it is indeed possible to balance, within a certain range, the achieved fps and output compression rate tradeoff by adjusting the I to P frames ratio at run-time.

### 2.4.4.2 Effect of Number of ME Processes

The next experiment shows the effect of varying the number of ME processes on the application's performance. A 1374 frames CIF resolution video is used as input, in which a maximum of 11 ME processes can be utilized. The total elapsed time of the application's execution is measured, for different numbers of ME processes, ranging from one to 11. The number of cores used is equal to the number of ME processes plus one.

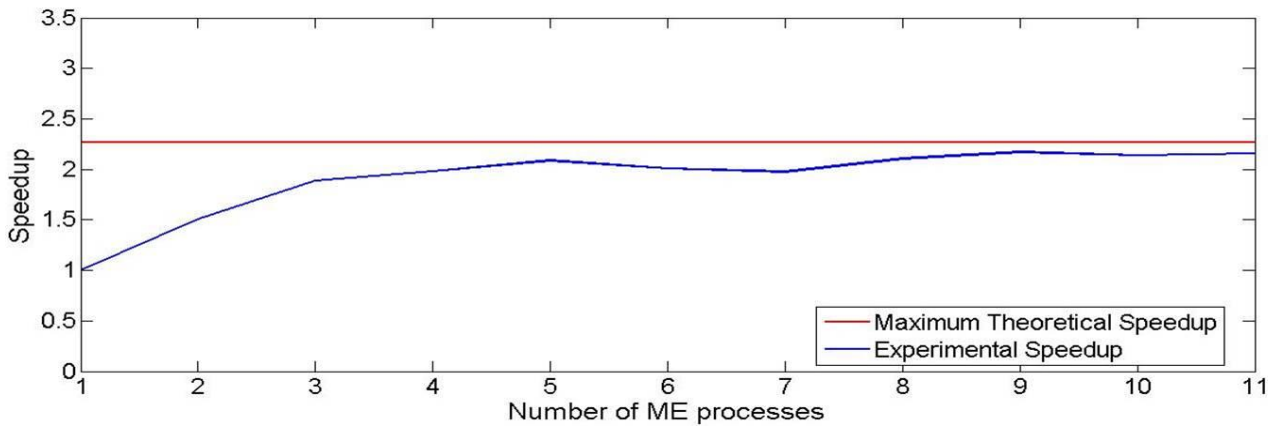

Figure 2-13. Speedup achieved for a varying number of ME processes, compared to the single ME process encoder.

Figure 2-13 shows the speedup achieved with each ME processes count, compared to the execution time of the application with one such process. It is evident that the speedup increases as the number of processes increases. This is because the predictions of more macroblocks are executed in parallel. As shown in the figure, the speedup is sub-linear. This is attributed to the fact that only a part of the application is parallelized, and even though some processes are pipelined, there are still parts of the encoder that are executed sequentially, i.e., the Deblock process always follows an Encode process. In addition, diminishing returns are witnessed as the number of processes increases, as is usually the case when increasing the parallelism of an application. This phenomenon can be attributed to the fact that the processes share the same system resources, such as the L2 cache memory and the memory bus, which can become congested as the communication traffic increases.

Nevertheless, using the maximum number of processes reaches close to the maximum achievable speedup. The divergence between the maximum theoretical and experimental speedups is attributed to the fact that the theoretical model does not incorporate some aspects of real-world systems, such as context switches and waiting time for memory accesses.

In addition, from the figure one can see that the execution with five ME processes achieves a speedup very close to the one achieved with 11 processes. Therefore, using five processes gives a more efficient encoder, as half the cores are used.

## 2.5 References - "Benchmarks for Embedded Systems"

1. **L. Schor, I. Bacivarov, D. Rai, H. Yang, S.-H. Kang, and L. Thiele**. Scenario-Based Design Flow for Mapping Streaming Applications onto On-Chip Many-Core Systems. Proc. Int'l Conf. on Compilers Architecture and Synthesis for Embedded Systems (CASES), pp. 71–80, 2012.

# 3. Brain Simulation Benchmark: measurements on the MPI version of DPSNN-STDP

The simulation of the activity and plasticity of large networks of neurons and synapses constitutes an extreme challenge for High Performance Computing systems. At smaller scales, it could be applied to embedded systems, if included in robots and medical systems. Indeed, the human brain learns through the dynamic of more than $10^{15}$ synapses and involves in its computations approximately $10^{11}$ neurons. The brain consumes a few tens of Watts; the problem is that, even at the highest levels of abstraction, a whole brain simulation would require tens of Giga Watts on computing systems designed using current technologies. Fault-tolerance and scaling to the required computational power are also overwhelming on conventional supercomputing architectures.

A neural network is composed of nodes (neurons). Each neuron is equipped with individual axonal and dendritic arborization. At the end of the arborization there are the synapses, connecting the neurons in a sparse network of nodes. The simulation of the brain activity can be performed at different level of abstractions. Actually, each neuron and each synapse is a complex biochemical engine. At the highest level of abstraction, the status of each synapse and neuron is represented by a few numerical values (e.g. the strength of the synapse connection, the time delay introduced by each synapse, the currents entering the neuron and the electric potential of the neuron membrane). According to this simplified model, spiking neural networks (SNN) and synaptic spike-timing dependent plasticity (STDP) should represent the computation and memory creation inside the brain.

In the framework of the EURETILE project, we are developing a mini-application benchmark of a Distributed Simulation of Polychronous Spiking Neural Network with synaptic Spike-Timing Dependent Plasticity (DPSNN-STDP), that should satisfy four requisites: 1- it should include the main challenges imposed by the brain simulation; 2- its coding complexity should be manageable; 3- it should be efficiently executed on standard distributed/parallel platforms, to profile the workload and the bottlenecks; 4- it should be compiled using the innovative EURETILE programming environment and executed on the EURETILE experimental execution platforms. The aim is to identify the bottlenecks of current software and hardware technologies, and to drive the development of suitable future architectures.

The finalization of an efficient and scalable version of the cortical simulation benchmark (DPSNN-STDP) coded in C++ plus MPI, and the measurement of its performance on a QUonG platform was one of the key goals of 2013 activities. Specifically, the objective of having an efficient and scaling reference code in MPI was to pose sound grounds for the final porting and measurements on the EURETILE environment, to be performed during 2014. The version of the DPSNN-STDP delivered at the end of 2013, described by this section of the document, is highly efficient and scalable, and support the description of a cortical area based on a bidimensional grid of cortical columns. In July 2013, a basic version of the complete code has been ported to DAL and delivered to the consortium for compilation and execution on the EURETILE software environment and execution platforms. During 2014, the highly optimized version described by this document will be ported to DAL for final measurements on the EURETILE platforms.

Project: **EURETILE** – European Reference Tiled Architecture Experiment
Grant Agreement no.: **247846**
Call: FP7-ICT-2009-4 Objective: FET - ICT-2009.8.1 Concurrent Tera-device Computing

## 3.1 Summary

We introduced[1] a natively distributed mini-application benchmark representative of plastic spiking neural network simulators. It can be used to measure performances of existing computing platforms and to drive the development of future parallel/distributed computing systems dedicated to the simulation of plastic spiking networks. The mini-application is designed to generate spiking behaviors and synaptic connectivity that do not change when the number of hardware processing nodes is varied, simplifying the quantitative study of scalability on commodity and custom architectures. Here, we present a first set of strong and weak scaling measures of DPSNN-STDP benchmark (Distributed Simulation of Polychronous Spiking Neural Network with synaptic Spike-Timing Dependent Plasticity). In this first test, we used the benchmark to exercise a small-scale cluster of commodity processors (varying the number of used physical cores from 1 to 128). The cluster was interconnected through a commodity network. Bidimensional grids of columns composed of Izhikevich neurons projected synapses locally and toward first, second and third neighboring columns. The size of the simulated network varied from 6.6 Giga synapses down to 200 K synapses. The code demonstrated to be fast and scalable: 115 wall clock seconds are required to simulate one second of activity and plasticity of a network composed by 1.6 G synapses (8 M neurons spiking at a mean firing rate of 22.5 Hz) on 128 hardware cores running @ 2.4 GHz. The mini-application has been designed to be easily interfaced with standard and custom software and hardware communication interfaces. It has been designed from its foundation to be natively distributed and parallel, and should not pose major obstacles against distribution and parallelization on several platforms. Here, we report about the relative weight of the computational and inter-process multicast blocks composing the application, and discuss the strong and weak scaling behavior. During 2014, we will further enhance it to enable the description of larger networks, more complex connectomes, and prepare it for distribution to a larger community. The DPSNN-STDP mini-application benchmark is developed in the framework of the EURETILE FET FP7 European project, but we acknowledge a useful exchange of ideas and goals with Paolo Del Giudice and Maurizio Mattia (CORTICON FET FP7 project).[2]

## 3.2 Introduction

Brain simulation is: 1- a scientific grand-challenge; 2- a source of requirements and architectural inspiration for future parallel/distributed computing systems, 3- a parallel/distributed coding challenge. The main focus of several neural network simulation projects is the search for a)-biological correctness; b)-flexibility in biological modeling; c)-scalability using commodity technology [e.g., NEURON (Carnevale, 2006, 2013); GENESIS (1988, 2013); NEST (Gewaltig, 2007);]. A second research line focuses more explicitly on computational challenges when running on commodity systems, with varying degrees of association to specific platforms echo-systems [e.g., Del Giudice, 2000; Modha, 2011; Izhikevich, 2008, Nageswaran, 2009]. An alternative research pathway is the development of specialized hardware, with varying degrees of flexibility allowed [e.g. SPINNAKER (Furber, 2012), SyNAPSE or BlueBrain projects]. Since 1984, the focus of our APE lab at INFN is the design and deployment of parallel/distributed architectures dedicated to numerical simulations (e.g. Avico et al., 1986; Paolucci, 1995). The present center of interest of APE lab is the development of custom interconnection networks (R Ammendola et al ., 2011). Indeed, the original purpose of the DPSNN-STDP project is the development of the simplest yet representative benchmark (i.e. a mini-application), to be used as a tool to characterize software and hardware architectures dedicated to neural simulations, and to drive the development of future generation simulation systems. Coded as a network of C++ processes, it is designed to be easily interfaced to both MPI and other (custom) Software/Hardware Communication Interfaces. The DPSNN-STDP mini-application benchmark has been designed to be natively distributed and parallel, and should not pose obstacles against distribution and parallelization on several competing platforms. It should capture major key features needed by large cortical simulations, should serve the community of developers of dedicated computing systems and should facilitate the tuning of commodity platforms. Moreover, the code demonstrated to be fast and scalable. For example, 115 wall clock seconds are required to simulate one second of activity and plasticity of a network composed by 1.6 G synapses (8 M neurons spiking at a mean firing rate of 22.5 Hz) on 128 hardware cores running @ 2.4 GHz. One of the explicit objectives is to maintain the code readable and its size at a minimum, to facilitate its usage as a benchmark. During 2014, we will further enhance it: 1- to enable the description of more complex connectomes; 2- to support other neural/synaptic models; 3- to prepare it for a possible distribution to a larger community of users.

---

[1] This section of the report is based on the results presented in: P.S. Paolucci et al., "Distributed simulation of polychronous and plastic spiking neural networks: strong and weak scaling of a representative mini-application benchmark executed on a small-scale commodity cluster", arXiv:1310.8478 [cs.DC]
[2] The CORTICONIC project is funded through the FET FP7 Grant Agreement no. 600806.

This chapter presents: 1- the measures of the strong and weak scaling behavior of the current svn release 801 of the DPSNN-STDP mini-application benchmark (developed during 2013), run on a small scale cluster of commodity processors interconnected through a commodity network; 2- the original hints for scalability derived from the profiling of the execution, that we think could be generalized to several simulators. The "Methods" section provides a compact description of the features of the current release of the DPSNN-STDP mini-application benchmark. The "Results" section reports the strong and weak scaling measures produced running the code on a small scale commodity cluster. We analyze the results and the hints for future developments in the sections "Discussion" and "Conclusions".

## 3.3 Methods

Here, we provide a compact summary about the internal structure of the current release of the DPSNN-STDP mini-application benchmark.

### 3.3.1 Each process describes a cluster of neurons and their incoming synapses

The full neural system is described by a network of C++ processes equipped with a message passing interface, agnostic of the specific message passing library. The full network is divided into clusters of neurons and their set of incoming synapses. The data structure that describes the synapse includes the information about the total transmission delay introduced by the axonal arborization that reaches it. The list of local synapses is further divided in sets according to the value of the axo-synaptic delay. Each C++ process describes and simulates a cluster of neurons and incoming synapses. The messages travelling between processes are sets of "axonal spikes" (i.e. they carry info about the identity of neurons that spiked and the original emission time of each spike). Axonal spikes are sent only toward those C++ processes where at least a target synapse exists for the axon. The knowledge of the original emission time of each spike and of the transmission delay introduced by each synapse allows for the management of synaptic STDP (Spike Timing Dependent Plasticity) (Song, 2000), which produces effects of Long Term Potentiation/Depression (LTP/LTD) of the synapses. This temporal information produces phenomena related to difference among delays along (chains of) individual synaptic arborizations (polychronism, see Izhikevich, 2006).

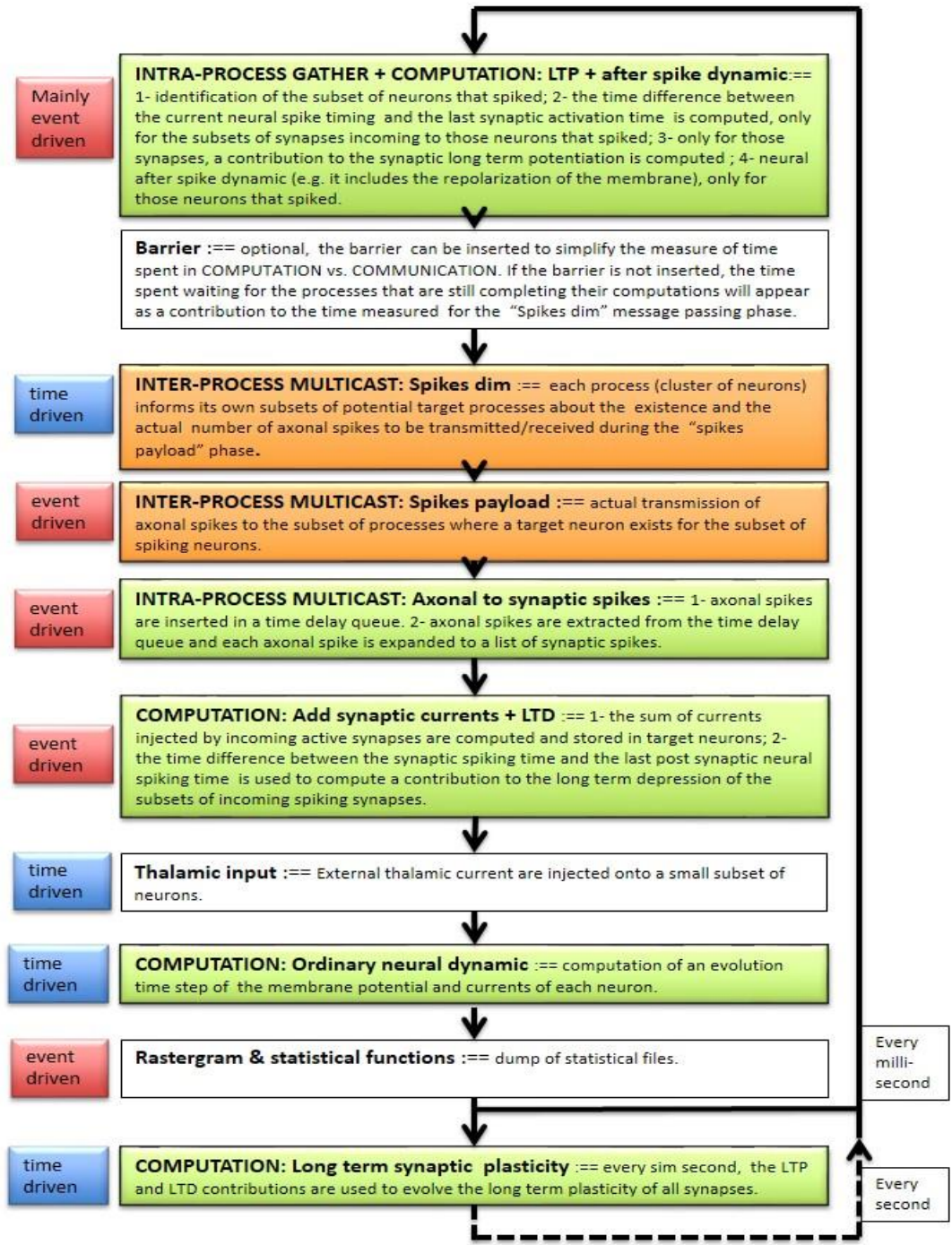

Figure 3-1. Cluster of neurons and incoming synapses are assigned to software process in a distributed simulation. Each process iterates over the blocks represented in this picture, that simulate the dynamic of neurons, the spiking and plasticity of synapses and the exchange of messages through axo-dendritic arborization. It is a flow of event-driven and time-driven computational and communication (inter-process and intra-process multicast) blocks. The measure of the (relative) execution times of the blocks guided the effort dedicated to the optimization of each block and can drive the development of dedicated hardware architecture.

### 3.3.2 Mixed time-driven and event driven simulation

There are two phases in a DPSNN-STDP simulation: 1- the creation of the initial state and structure of the neural network; 2- followed by the simulation of the dynamic of neurons and synapses. The initial construction of the system (1) includes the creation of the network of axonal polychronous arborizations and synapses that interconnect the system.
We adopted a combined event-driven and time-driven approach (Morrison et al, 2005) for the simulation of the neural and synaptic dynamic:
- *Event-driven simulation*, for synaptic dynamics.
- *Time-driven simulation*, for neural dynamics.

The phase of simulation of the dynamic (2) can be further decomposed into an iteration over the following steps: 2.1- the subset of neurons that produced spikes during the previous time-driven simulation step induces an event-driven long term potentiation of their incoming synapses using the STDP scheme and perform other post-spiking activities; 2.2 - spikes are sent through axonal arborizations to the cluster of neurons where target synapses exist; 2.3 - axonal spikes delivered to a process are queued into a list, for usage during this time-step and a window of subsequent time steps; 2.4 - axonal spikes, classified according to their time of arrival, are delivered through local axonal arborizations to the subset of their target synapses of appropriate delay; 2.5 - synapses inject currents into their target neuron; 2.6 - target neurons induce an event-driven long term depression on the subset of synapses that injected currents; 2.7 - the full set of neurons perform one step of time evolution.

### 3.3.3 Spiking neuron model

Hybrid models describe the continuous evolution of several state variables (including a "membrane voltage" and auxiliary "currents") and discrete events associated to the spiking event, i.e. special rules applied to (a subset of) the state variables. Well known are the Hodgkin-Huxley (HH) (Huxley, 1952), the leaky integrate-and-fire (LIF) and the Izhikevich (IZH) (Izhikevich, 2003). For this experiment we adopted the IZH model which is computationally efficient (13 – 26 operations per simulated ms per neuron), and yet capable of replicating the spiking behaviour of several neuron types (Izhikevich, 2004).

$$\begin{cases} \boldsymbol{if\ v(t)} < v_{peak} \quad then \quad \begin{cases} \frac{\Delta \boldsymbol{v}}{\Delta \boldsymbol{t}} = v(t)^2 - u(t) + I(t) \\ \frac{\Delta \boldsymbol{u}}{\Delta \boldsymbol{t}} = a(bv(t) - c) \end{cases} \\ \boldsymbol{if\ v(t)} \geq v_{peak} \quad then \quad \begin{cases} \boldsymbol{v(t)} = v_{peak} \\ \boldsymbol{v(t + \Delta t)} \leftarrow \boldsymbol{c} \\ \boldsymbol{u(t + \Delta t)} \leftarrow \boldsymbol{u(t)} + \boldsymbol{d} \end{cases} \end{cases}$$

where:
- *v* (t) is the neural membrane potential. This is the key observable; we say that when *v* reaches **v_peak** a "neural spike" happened;
- I(t) is the potential change generated by the sum of all synapses incoming to the neuron. Incoming currents are present if spikes arrived form presynaptic neurons;
- u(t) is an auxiliary variable (the recovery current bringing back v to equilibrium);
- a, b, c, d are four parameters, constant for each neuron kind, by varying them the same equation models several kind of known neural types.

In this experiment we used a mix of 80% excitatory RS Izhikevich neurons (i.e.: a=0.02, b=0.2, c = -65.0 mV, d=8.0) and 20% inhibitory FS neurons (obtained by setting a=0.1, b=0.2, c = -65.0 mV, d=2.0). **V$_{peak}$** was set at 30 mV.

### 3.3.4   Synaptic update: spike-timing dependent plasticity

Let us define t = t$_{post}$ − t$_{pre}$ − d$_{axon}$, the time difference between the post-synaptic spike time, and the time of arrival of a spike originated by a presynaptic neuron at an original emission time t$_{pre}$, that arrives at the target after an axonal delay d$_{axon}$. We implemented the following STDP rule, to compute the ΔW$_{pre,post}$ change to the synaptic strength (Song et a., 2000). A$_+$, A$_-$, τ$_+$, τ$_-$, are parameters which permits to match the model on different types of neurons and biochemical contexts.

$$t = t_{post} - t_{pre} - d_{axon} \quad \begin{cases} if \ t \geq 0 & \Delta W_{pre,post} = A_+ e^{-\frac{t}{\tau+}} \\ if \ t < 0 & \Delta W_{pre,post} = A_- e^{\frac{t}{\tau-}} \end{cases}$$

The synapse is maximally potentiated if the delay introduced by the axon carries the signal to the target just before the post-synaptic spike (i.e. it is probably the cause of the spike). The synapse is maximally depressed if the signal arrives just late.

### 3.3.5   Projection of synapses with individual delay

In our polychronous networks each neuron *i* = 1..N projects its set of forward synapses *j*=1..M, each one characterized by its individual delay *D$^{i,j}$*, plastic weight *W$^{i,j}$* and target neuron K$^{i,j}$. In this set of tests, M was fixed to 200 for all neurons. Inhibitory neurons projected synapses only toward excitatory neurons located in the same column. Instead, excitatory neurons projected also to neighboring columns, as discussed in a following section. For this experiment, we assigned delays in the range between 1 and 20 ms. Inhibitory synapses were assigned with the minimum delay, while excitatory delays were assigned with a uniform distribution of delays. If a neuron fires at a time *t$_i$*, each forward synapse will have to inject a current *W$^{i,j}$* at a synaptic specific time *t$_i$* + *D$^{i,j}$*. The current *W$^{i,j}$* will be injected in the target neuron *k$_j$*=K$^{i,j}$, where it will add to the currents arriving in the same time step from other source neurons, to form the total external incoming current *I$_k$(t)* which contributes to the dynamic of neuron *k$_j$*.

### 3.3.6   Bidimensional arrays of neural columns and their distribution on the processes and processors

In this experiment of strong and weak scaling behavior, we arranged the neurons in "columns", each one composed by one thousand neurons. Columns are then arranged in bidimensional grids (see Figure 3-2). Each excitatory neuron projected 76% of its synapses to neurons in its own column, 12% of its synapse toward neurons in first neighboring columns, 8% towards second neighboring columns and 4% toward neurons in third neighboring columns (see picture). Inhibitory neurons projected their synapses only towards excitatory neurons in the same column. We varied the total number of columns between one and 32768 (a 256x128 grid of neural columns, for a total of 32.7 million neuron). Each process can either host a fraction of a column (e.g. 1/8, ¼ …), a whole single column, or several columns (in the present version, up to 32 columns per process). When performing strong and weak scaling measures, periodic boundary conditions are used for columns at the boundary of the grids, to get more homogeneous spiking rates for different numbers of columns. For grids so small not to have enough distinct neighbors, the periodic boundary rule can end projecting more synapses on the same target column than expected for a large grid. Actually, in the case of a single column, all synapses are projected by the column to itself.

A sample grid of 64=8x8 neural columns.

Excitatory neurons projects 76% of their synapses toward neurons located in the same column, 3% to first neighboring columns, 2% to second neighbours and 1% to third neighbours.

Strong scaling measures: a,b,c) are examples of distribution of a grid composed of 64 neural columns over a varying number of software processes and computational cores.

One computational core can host one or more software processes

The number of software processes assigned to each computational core has been changed during the strong scaling measure.

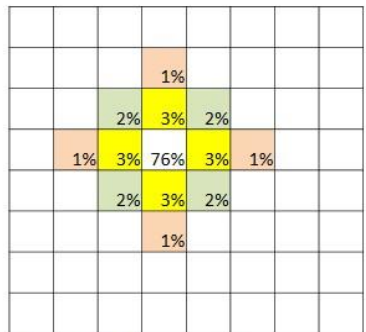

a) 64 neural columns on 64 sw processes - The number of software processes is equal to the number of neural columns.

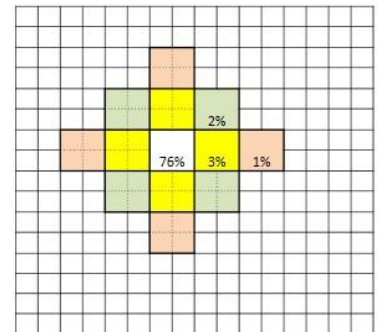

b) 64 neural columns on 256 sw processes - Each neural column is distributed among four software processes.

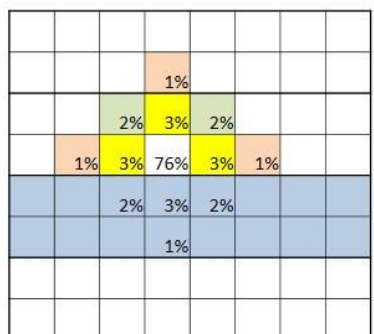

c) 64 neural columns on 4 sw processes - In this case, 16 neural columns are managed by each software process.

The simulation of the same grid of neural columns produces identical results on all distributions.

Figure 3-2. Example of distribution of an identical problem over a varying number of software processes and computational cores. In this figure, the grid to be distributed is composed by 8x8=64 neural columns. The DPSNN-STDP simulator produces the same external stimulus, synaptic structure and spiking activity on all distributions. We varied the number of columns (from 1 to 32768), the number of software processes (from 1 to 256) and the number of hardware cores (from 1 to 128) along the strong and weak scaling measures described in this paper.

### 3.3.7 Distributed generation of reproducible connections and external "thalamic" stimulus

We mention a feature that has been of some importance to simplify the execution of repeatable strong and weak scaling measures, while varying the number of processes and hardware resources (e.g. processors). We mean, the capability to initialize in a distributed manner an identical network and provide, again in a distributed manner, the same external "thalamic" stimulus to a network composed by a given grid of neural columns, distributed over a varying number of software processes and hardware processors. In a system with N total neurons, distributed among H software processes we can assign a fair share of locN = N/H neurons per software process, and the global and local identities of neurons can be easily computed using the local identifiers of processes and neurons. If there is a grid of CFT = CFX x CFY neural columns, and this info is known to each process, it will be possible for each process to generate forward connectivity patterns that does not depend on the number of processes/hardware processors. The same can be done to generate patterns of external "thalamic" stimulus to the network, e.g. prescribing the number of events per ms per neural column.

### 3.3.8 Production of Observables. The behavior of individual neurons.

The DPSNN-STDP code can produce files tracing several observables (list of individual spiking times and spiking neuron identity, mean spiking rates, membrane potentials, synaptic values). The code is equipped with facilities for distributed measurement of time spent on execution of individual routines and sections of the code based on the `MPI_WTIME` function.

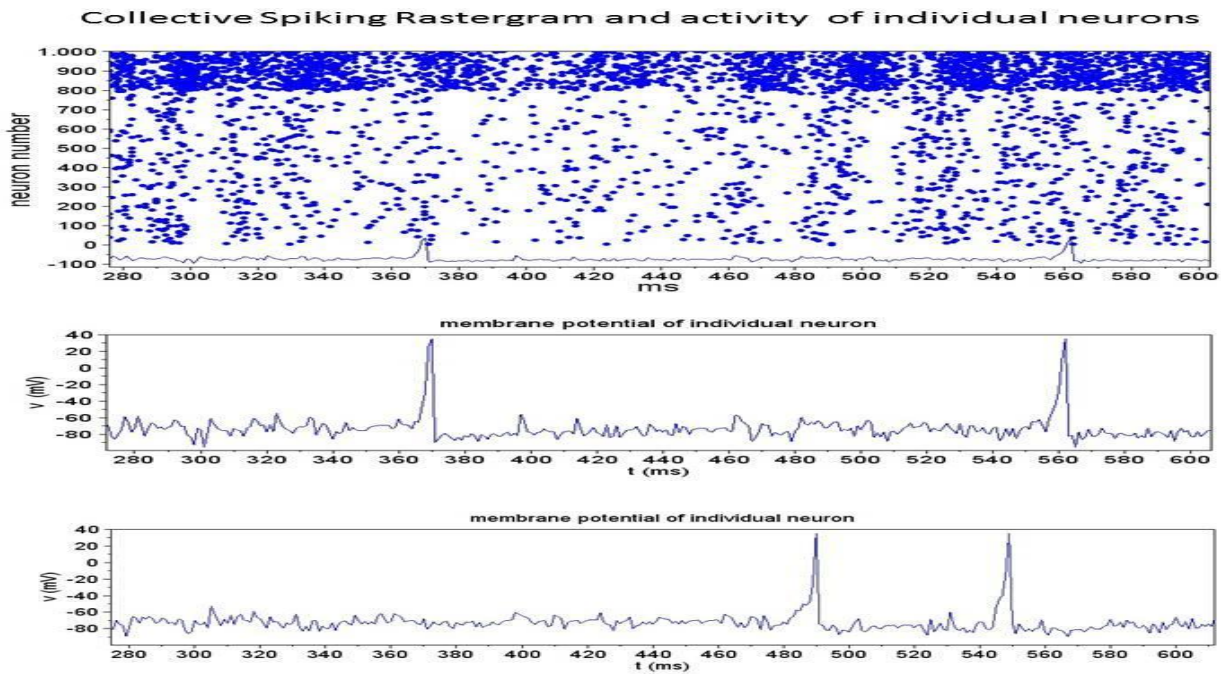

Figure 3-3 A sample trace of 320 ms of spiking activity, produced by the DPSNN-STDP code. In this case, it is the simulation of a single neural column, composed by 1000 neurons (80% excitatory RS, 20% inhibitory FS Izhikevich neurons). Above, each dot in the rastergram represents a spiking event. Below, the traces of the membrane potential of two excitatory neurons.

If necessary, the trace of the evolution of the membrane potential and other state variables of individual neurons can be activated.

The membrane potential of a "resting" neuron fluctuates around a -70mV potential as a result of its own activities and of the perturbation produced by signals produced by other neurons. When the neuron decides to "fire", its membrane potential (v state variable) starts to climb to positive voltages. If a "spike" happened, the Izhikevich rules drops the voltage down to its after spike potential, but the internal variable u keeps a memory of the past. This event is the "spike" of an individual neuron, which is propagated by the axon, and reaches a set of synapses after a time delay, specific for each synapse, which depends on the distance travelled. When reached by the spike, each synapse produces a perturbation of the membrane potential of the target, which depend on its "strength" (here, we consider the simplistic case of the neural soma and dendritic arborization represented by a single "compartment", i.e. a single u(t)-v(t) pair in the case of the Izhikevich equation).

The picture reports about 320 ms of collective spiking activity of a single column of 1000 neurons (800 RS excitatory, 200 FS inhibitory), and the evolution of the individual membrane potential of two neurons. In the "rastergram" the horizontal axis is the simulation time, the vertical axis the identifier of individual neurons. Each dot in the rastergram represents a spiking event.

### 3.3.9 Representation of spiking messages

Spiking messages are sent using an address event representation (AER): we send "axonal spike" messages that carry the identifiers of spiking neurons and are packed in groups that have the same spike emission time and the same target process (i.e. same target cluster of neurons). Our strategy is to defer as much as possible the arborization of the "axon", to reduce the load on the network and unnecessary wait barrier (i.e. waiting for the completion of computations of cluster of neurons from which a process does not expect messages). To this purpose, we perform some preparatory actions during the network initialization phase (performed once at the beginning of the simulation), to reduce the number of active communication channels during the iterative simulation phase.

### 3.3.10 Initial construction of the connectivity infrastructure

During the initialization phase, each process can contribute to create the awareness about the subset of processes that should be listened to, during next simulation iterations. At the end of this construction phase, each "target" process should know about the subset of "source" processes that need to communicate with it, and should have created its database of locally incoming axons and synapses. A simple implementation of the construction phase can be realized using two steps.

During the first step, each source process informs other processes about the existence of incoming axons and about the number of incoming synapses to be established. A single word, the synapse counter, is communicated among pairs of processes. Under MPI, this can be achieved by an `MPI_Alltoall()`. Performed once, and with a single word payload, the cost of this first step, creates a cumulative network load proportional to the square of the number of processes. The cost of this operation is negligible in the range of processes explored by this paper.

The second step transfers the identities of synapses to be created on each target process. Under MPI, the payload, a list of synapses specific for each pair in the subset of processes to be connected, can be transferred using a call to the `MPI_alltoallv()` library function. The cumulative load created by this second step is proportional to the product between the total number of processes and the subset of target processes reached by each source process.

The first step produces two effects: 1- it reduces the cost of initial construction of synapses, second step of the construction phase; 2- the knowledge about the non existence of a connection between a pair of processes can be used to reduce the cost of spiking transmission during the simulation iterations.

### 3.3.11 Delivery of spiking messages during the simulation phase

Here, we describe the present implementation of the delivery of spiking messages. In this first implementation, we did not take advantage of the possibility of delivering spikes to targets just before the deadline imposed by the synaptic specific delay. Instead, we used a synchronous approach: all spikes are delivered to target processes before proceeding to the simulation of the next time iteration of the neural dynamic.

The delivery of spiking messages can be split in two steps, with communications directed toward subsets of decreasing sizes.

During the first step, single word messages (spike counters) are sent to the subset of potentially connected target processes. On each pair of source-target process subset, the individual spike counter informs about the actual payload (i.e. axonal spikes) that will have to be delivered, or about the absence of spikes to be transmitted between the pair. The knowledge of the subset has been created during the first step of the initialization phase, described in a previous section.

The second step uses the spiking counter info to establish a communication channel only between pairs of processes that actually need to transfer an axonal spikes payload during the current simulation time iteration.

On MPI, both steps can be implemented using calls to the `MPI_Alltoallv()` library function. However the two calls establish actual channels among sets of processes of decreasing size, as described just above.

For the simple bidimensional grid of neural columns and for the mapping on processes used in this experiment this implementation demonstrated to be quite efficient, as reported by the measures presented in the "Results" section, further refined in the "Discussion" section. However, we expect that the delivery of spiking messages will be one of the key point still to be optimized when white area "connectomes" will be introduced, describing the communication channels among a multiplicity of remote cortical areas.

## 3.4 Results

This section presents the strong and weak scaling behavior of the current revision of the DPSNN-STDP mini-app benchmark, run on a small scale commodity cluster.

We run on a cluster of sixteen dual socket quad core servers, interconnected through a 40 Gb/s commodity network, for a maximum of 128 physical cores running 2.4 GHz.[3]. Each physical core supported two simultaneous threads.

During this experiment, for each neural network size, we checked that the list of spiking neurons and their timings were identical for all run performed using a variable number of software processes and/or physical cores.

For all configurations used for the scaling measures, each neurons projected 200 forward synapses. Neurons were always grouped in "columns", each column composed of 1000 neurons (80% excitatory, 20% inhibitory). As described in a previous section, neural columns were organized in bidimensional grids, and each excitatory neuron projected part of its synapses (76%) toward neurons, and the remainder was distributed among first, second and third neighboring columns, with decreasing proportions (3% toward each of the 4 first neighbouring columns, 2% toward each second neighbour, 1% toward each third neighbour). We varied the size of the bidimensional grid of columns from a minimum configuration of 1x1 to a maximum of 256x128 neural columns. This way, we obtained configurations with a varying number of synapses: from a minimum of 200 K synapses to a maximum of 6.6 G synapses. Each network was distributed on a variable number of software processes, and then assigned to a variable number of physical cores. Each MPI process hosted a maximum of 1024 neural columns (i.e. 1024 K neurons and 51.2 M synapses per process), and a minimum of 1/8 of neural column (i.e. 125 neurons and 5 K synapses per process).

---

[3]Each server is a 1U SuperMicro X8DTG-D. Each node in the cluster is a dual socket. Each socket hosts one quad-core Intel(R) Xeon(R) CPU E5620 (max clock @ 2.40GHz). On each core HyperThreading is enabled (two threads per core). Each node is equipped with a Mellanox InfiniBand board, the MT26428 [ConnectX VPI PCIe 2.0 5GT/s - IB QDR (40Gb/s data rate)]. 16 nodes are connected using a Mellanox Switch.

Project: **EURETILE** – European Reference Tiled Architecture Experiment
Grant Agreement no.: **247846**
Call: FP7-ICT-2009-4 Objective: FET - ICT-2009.8.1 Concurrent Tera-device Computing

Table 3-1. We report in the table a subset of the measures used to produce the strong and weak scaling graphs reported in a following section. In particular the table reports: 1- a subset of the configurations and 2- a subset of the measured execution times. We run different problem sizes, from 200 K synapses to 6.6 billion synapses. Each neural network size (a column in the table) was distributed using a varying number of MPI processes, and run on a varying number of physical computational resources. The simulation of a given network size produced an identical spiking and plasticity behavior (e.g. firing activity) over 2000 ms of simulated activity, for all distributions among software processes and/or hardware cores.

| Total synapses | 200 K | 800 K | 3.2 M | 12.8 M | 51.2 M | 204.8 M | 819.2 M | 3.2 G | 6.6 G |
|---|---|---|---|---|---|---|---|---|---|
| Total neurons | 1 K | 4 K | 16 K | 64 K | 256 K | 1024 K | 4.096 M | 16.4 M | 32.8 M |
| Grid of neural columns | 1 x 1 | 2 x 2 | 4 x 4 | 8 x 8 | 16 x16 | 32 x 32 | 64 x 64 | 128x128 | 256x128 |
| Mean firing rate (Hz) | 27 | 24 | 26 | 23 | 22 | 23 | 20 | 22 | 19 |
| Used cores[4] (min-max) | 1-8 | 1-32 | 1-128 | 1-128 | 1-128 | 1-128 | 4-128 | 64-128 | 64-128 |
| MPI processes | 1-8 | 1-32 | 1-128 | 1–256 | 1-256 | 1-256 | 4-256 | 64-256 | 128 |
| Execution time[5] (execution sec / simulated sec) | 0.15 | 0.4 | 1.80 | 3.05 | 6.85 | 20.0 | 59 | 211 | 386 |
| Normalized execution time[6]: execution time / (firing rate × total syn × simulated second) | $2.73 \times 10^{-8}$ | $5.36 \times 10^{-9}$ | $2.41 \times 10^{-8}$ | $4.22 \times 10^{-9}$ | $6.0 \times 10^{-9}$ | $4.22 \times 10^{-9}$ | $3.61 \times 10^{-9}$ | $2.94 \times 10^{-9}$ | $3.07 \times 10^{-9}$ |

Due to the hardware support for two simultaneous threads per core, we observed that, for larger networks (i.e. when more than 16K neurons can be allocated on each process), the best execution time was reached when two MPI software processes were launched on each physical core.

The time step of the simulation was set to 1 ms. We measured the execution time needed to simulate 2 seconds of spiking activity and synaptic plasticity (i.e. 2000 simulation iterations, using a ms time step). Specifically, we measured the wall clock needed to simulate the 8$^{th}$ and 9$^{th}$ second of evolution (i.e. we left the system evolve for 8000 simulation steps, before taking our measure). During the 2 seconds selected for the measures, the networks exhibited mean firing rates in the range between 19 and 27 Hz.

The distributed DPSNN-STDP code demonstrated to be competitively fast. For example, only 115 wall clock seconds are required to simulate one second of activity and plasticity of a network composed by 1.6 G synapses (8 M neurons spiking at a mean firing rate of 22.5 Hz) on 128 hardware cores running @ 2.4 GHz. Looking for further improvements and for possible showstoppers when scaling to larger configurations, we profiled the relative weight of the computational and inter-process multicast blocks composing the application, and analyzed the strong and weak scaling behavior, as discussed in the following sections.

---

[4] Each cores @2.4 Ghz part of a quad-core Intel(R) Xeon(R) E5620.
[5] Using the "max" number of cores reported in this table
[6] See the "Strong scaling" section for a discussion about the unit of measure for the normalized execution speed.

### 3.4.1 Strong scaling

The computational load needed to solve the simulation problem is expected to be proportional: a) to the number of simulated synapses, b) to the firing rate and c) to the physical time to be simulated. Therefore, if we divided the measured execution time on a given number of computational cores by the product of a, b and c, we should obtain an estimate of the execution time needed per synapse per second. In first approximation, we could expect this number to be similar for different problem sizes. Then, an ideal code distributed on an ideal machine should half its execution time when doubling the number of computational cores assigned to the solution of the problem. Here below a picture of the scaling we observed running the DPSNN-STDP code. However, in Figure 3-4 we observe a scaling on the log-log graph, which, is still not ideal. For example, let us discuss the measures of the 204 M synapse case. On the configuration with 128 hardware cores, 20 seconds were needed to simulate one seconds of spiking and plasticity at a mean firing rate of 23 Hz; while 828 second were needed to produce the same results on a single core. In Figure 3-4, this corresponds to a reduction from $1.75 \times 10^{-7}$ s to $4.22 \times 10^{-9}$ s of the wall clock time needed per synapse, normalized to the firing rate. Instead of decreasing by a factor 128, the actual speed-up is only 41.5. At each doubling in the number of cores, the actual speed-up is only 1.7. Using this slope as representative, the multiplication of the number of cores by a factor 128 would decrease the simulation time 3 times worst that an ideal scaling. We analyze the results in the "Discussion" section of this document.

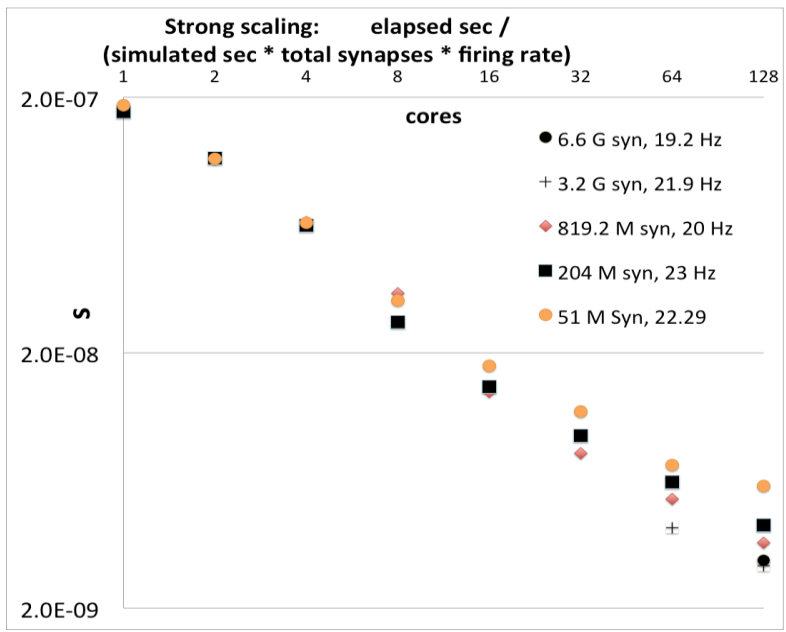

Figure 3-4 Strong scaling of the DPSNN-STDP mini-app benchmark. For an ideal scaling the execution time should grow proportionally to the number of synapses and to the firing rate, and should reduce proportionally to the number of cores applied to the execution.

### 3.4.2 Weak scaling

If, instead of dividing the execution times by the total number of synapses, we divided it by the number of synapses assigned to each computational core, we should obtain, for the same consideration of the previous section, a value that for an ideal code, executing on an ideal machine should be constant for different network sizes and number of computational cores assigned to the solution of the problem.
Figure 3-5 is the graph of our measures. Looking for hints about how to further enhance the scalability features of the system, we observe an interesting feature: when the simulation is distributed among a large number of hardware cores, the simulation runs relatively faster for larger configurations (i.e. configurations that host more "neural columns" per hardware core/software process).

We analyze this point in the next "Discussion" section.

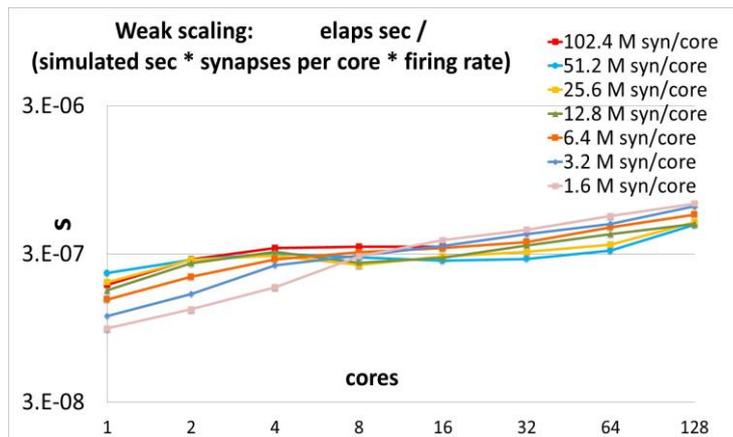

Figure 3-5 Weak scaling of the DPSNN-STDP benchmark

### 3.4.3 Profiling of the relative weight of the computational and inter-process multicast blocks

Table 3-2. Profiling of execution time of the DPSNN-STDP mini-application benchmark: relative weights of computation and communication blocks. The relative weight demonstrated to be quite stable for different network sizes and number of cores used for the simulation. Quite interesting is the weight of the (optional) synchronization barrier. If the barrier is removed, a greater weight would appear in the "Spikes dim" phase. The data used for the table here inserted are relative to a network of 100 M synapses executed by 64 MPI processes on 32 cores[7].

| Function of the block | Relative execution time | Note |
|---|---|---|
| Long term potentiation + after spike dynamic | (9.7 ±0.7)% | Gather[8] + computation |
| Barrier (optional)[9] | (29.9±6.1)% | Workload fluctuations |
| Communication: inter-process multicast: Spikes dim | (0.77±0.10)% | Message passing |
| Communication: inter-process multicast: Spikes payload | (0.82±0.20)% | Message passing |
| Axonal to synaptic spikes: intra-process multicast | (16.8±2.3)% | Dereferencing[10] |
| Add synaptic currents + long term depression | (19.2 ±2.7)% | Computation |
| Thalamic input[11] | 0.01% | Simplified model |
| Ordinary neural dynamic | (11.8±1.4)% | Computation |
| Rastergram & other statistical functions | (1.9±0.1)% | Computation |
| Long term synaptic plasticity | (9.2±1.8)% | Computation |

---

[7] See the notes to Table 3-1 for what concerns the hardware configuration used for the measures.

[8] The benchmarked software implementation is based on sparse accesses from the target neuron to the global list of incoming synapses. In a hardware implementation, based on several independent memory banks, if all synapses incoming to the same neuron were stored in contiguity, this task could be easily accelerated.

[9] If the barrier is not inserted, the time spent waiting for the processes that are still completing their computations appear as a contribution to the time measured for the "spikes dim" communication phase. We verified that the deviations from ideal strong scaling, can be entirely attributed to the cost of measured fluctuations in workload execution (represented by the Barrier block) and to (a very small) cost of communications.

[10] The task to be performed is an "intra-process" multicast, from axons to specific lists of synapses. Instead, the benchmarked software implementation is based on two levels of dereferencing.

[11] In this simulation the thalamic input is computed using a simple statistical model. Actually, this is one of the interface between the neural network and the "external" world, so its weight would greatly increase and add to that of other interfaces to be added.

As discussed in the "Methods" section, each process iterates over a few main computation and inter-process multicast blocks, described in Figure 3-1. Here, in Table 3-1Table 3-2, we report the relative time spent on each block during execution, using the same division in functional blocks.

## 3.5 Discussion

### 3.5.1 Impact of unbalances in individual process workload and execution environment

As discussed in the "Strong scaling" section, a typical behavior of our code is the following: the execution time reduces by a factor 42, when running on 128 cores, compared to the time needed to execute the same configuration on a single core. Initially, we had the temptation to attribute the deviation from ideal scaling to the cost of communications, the usual suspect in the implementation of distributed applications. Therefore, we took great care in optimizing the communication algorithm. After this optimization, the improved communication mechanism described by the "Delivery of spiking messages during the simulation phase" section of this documents produces an expected low cost of communication for the simulated biological connectivity (a bidimensional grid of "cortical modules" with first, second and third neighbor communication). Indeed, the insertion of barriers, when measuring the relative time spent by each process in the set of computation and inter-process multicast blocks, demonstrated that the major cause for deviation from an ideal scaling is the fluctuation in the execution time of different processes, which force some processes to wait for others. Indeed, we verified that the deviations from ideal strong scaling, reported in Figure 3-4, can be entirely attributed to the cost of measured fluctuations in workload (represented by the Barrier block in Table 3-1) and to (small) cost of communications. There are a few main sources for the difference in execution times, including: 1- the workload of the individual process (i.e. the activity of a cluster of neurons and synapses) changes, depending on the local mean spiking rate and on the number of incoming spikes; 2- the execution time fluctuates even for fixed input and output spikes, due to the software execution environment and to the specificities of the individual hardware node whereon the process is executing. Therefore, we expect load-balancing to play an important role in the simulation of biologically realistic, large scale networks.

### 3.5.2 A quantitative basis to drive the development of specialized hardware and software

During the development and optimization of the code, we used the measure of the relative execution times (like those reported in Table 3-2) to drive both the definitions of the functional blocks and the time dedicated to their optimization. Hints from frequent measures of the strong and weak scaling behaviour (Figure 3-4 and Figure 3-5) has been also instrumental to the optimization of the code, in particular for the design of the inter-process communication. The data structures of each process and those inside each functional block, as well as the structure adopted for the execution of each block have been designed and coded in a manner that should simplify further distribution/parallelization of the activity of each block. Therefore, we consider the present release of the code a good starting point for the development of specialized hardware and software architectures.

### 3.5.3 From a bidimensional grid of cortical modules to a long-range, complex connectome

Here, we measured the performance of the DPSNN-STDP mini application benchmark exercised on a bidimensional grid of cortical modules. This interconnection is representative of the local structure, inside a cortical area. However, the simulation of long range "white-matter" connectomes, i.e. of the interconnection among cortical areas, will increase the number of target processes and the complexity of an efficient algorithm for spike delivery.

## 3.6 Conclusions

We introduced a natively distributed mini-application benchmark representative of plastic spiking neural network simulators. It can be used to measure performances of existing computing platforms and to drive the development of future parallel/distributed computing systems dedicated to the simulation of plastic spiking networks. The mini-application is designed to generate identical spiking behaviors and network topologies over a varying number of processing nodes, simplifying the quantitative study of scalability on commodity and custom architectures. Here, as a test case, we presented a first set of strong and weak scaling measures on a small scale cluster of commodity processors (varying the number of used physical cores and the problem size) and searched for the cause of the deviation from the ideal scaling behavior for a simple neural network structure (bidimensional grids of neural columns, connected to first, second and third neighboring columns).

The cause of this deviation seems to be the jitter in the execution time of each process. This can be due to two main factors: 1) an intrinsic difference between processes (they differ from each other, for example, in the local spiking rate), 2) some extrinsic differences (due to several factor such as the impact of the operating system, the local adaptation of the processor clock speed, etc.). This analysis is valid for the bidimensional topology of neural columns (and software processes). In a future work, we will study the impact of the support of more complex connectomes on the communication, which, in the present release, seems to be well optimized for scalability, at least for the explored size of neural networks and hardware resources.

More in general, our expectation is that the potential performance improvements from dedicated software and hardware co-design solutions will grow, when more complex interconnection topologies (e.g. inter-areal connectomes) will be simulated. The mini-application has been designed to be easily interfaced with standard and custom software and hardware communication interfaces and permit easy measurements of scalability. It has been designed from its foundation to be natively distributed and parallel, and should not pose major obstacles against distribution and parallelization on several platforms. During 2014, we will further enhance it to enable the description of larger networks, more complex connectomes, and prepare it for distribution to a larger community. Meanwhile, the DPSNN-STDP mini-application benchmark will be validated against biologically significant cases.

## 3.7 Acknowledgements

The work of the INFN team that developed the DPSNN-STDP benchmark has been supported by the EURETILE European integrated FP7 project grant no. 247846. However we also acknowledge the valuable exchanges of ideas with Paolo Del Giudice and Maurizio Mattia, supported by the FP7 project CORTICONIC (grant no. 600806).

## 3.8 References - "Brain Simulation Benchmark"